\documentclass[aps,amsmath,amssymb,preprint,prd,showpacs,showkeys]{revtex4-1}
\usepackage[b]{esvect}
\usepackage{graphicx}
\usepackage{dcolumn}

\usepackage{bm}
\usepackage{makecell}
\usepackage{footnote}
\usepackage{multirow}
\usepackage{xcolor}
\usepackage[colorlinks=true]{hyperref}
\usepackage{axodraw}
\usepackage{subcaption}
\begin{document}
\title{Study of semileptonic decay of $\bar{B}_s^0\to \phi l^+ l^-$ in QCD sum rule}

\author{Ying-Quan Peng}
\email{2120170119@mail.nankai.edu.cn}
\author{Mao-Zhi Yang}
\email{yangmz@nankai.edu.cn}
\affiliation{School of Physics, Nankai University,Tianjin 300071, People's Republic of China}

\begin{abstract}
In this work we study the semi-leptonic decay of $\bar{B}_s^0\to \phi l^+ l^-$ ($l=e, \mu, \tau$) with
QCD sum rule method. We calculate the $\bar{B}_s^0\to \phi$ translation form factors relevant to
this semi-leptonic decay, then the branching ratios of $\bar{B}_s^0\to \phi l^+ l^-$ ($l=e, \mu, \tau$)
decays are calculated with the form factors obtained here. Our result for the branching ratio of
$\bar{B}_s^0\to \phi\mu^+ \mu^-$ agrees very well with the recent experimental data. For the unmeasured
decay modes such as $\bar{B}_s^0\to \phi e^+ e^-$ and $\bar{B}_s^0\to \phi\tau^+ \tau^-$, we give
theoretical predictions.
\end{abstract}


\maketitle

\section{Introduction}
In the Standard Model, flavor-changing neutral current (FCNC) induced processes are forbidden at tree level. They
can only occur via loop diagrams. Meanwhile they are also sensitive to contributions
of new physics. Particles of new physics may contribute via loop diagrams as ``virtual particles",
thereby affecting the physical processes induced by FCNC.
With continuous improvement of experimental accuracy, FCNC processes play an increasingly important role in the new physics
research in heavy flavour physics. The most typical process is the one caused by $b\to sl^+ l^-$, such as
the rare semi-leptonic decays of $\bar{B}_s^0\to \phi l^+ l^-$ ($l=e, \mu, \tau$).

In the past two decades, the decays of $\bar{B}_s^0\to \phi l^+ l^-$ ($l=e, \mu, \tau$) have been studied by using several different approaches such as lattice QCD (LQCD) \cite{LQCD}, QCD light-cone sum rule (LCSR) \cite{LCSR,LCSR-improve}, constituent quark model (CQM) \cite{CQM-G,CQM-Me}, QCD sum rule \cite{R.K-SumR}, relativistic quark model (RQM) \cite{FG} and covariant quark model \cite{CoQM}. The method of QCD sum rule (SR) was originally developed by Shifman, Vainshtein and Zakharov in the late 1970s \cite{SVZ1,SVZ2}, which was then widely applied to the calculation of hadronic physics \cite{Colan2000}. Several years ago the translation form factors of $\Bar{B}^0_s\to\phi$ in $\bar{B}_s^0\to \phi l^+ l^-$ ($l=e, \mu, \tau$) decays were calculated with QCD sum rule in Ref. \cite{R.K-SumR}. Compared with other results, some form factors obtained in Ref. \cite{R.K-SumR} are different by negative signs, which are not simply due to different definition for the form factors.

Experimentally, LHCb Collaboration updated the measurement of the branching ratio of
$\bar{B}_s^0\to \phi \mu^+ \mu^-$ recently \cite{15AQ} ,
\begin{equation}\label{1}
\mbox{Br}(\bar{B}_s^0\to \phi \mu^+ \mu^-)=(7.97^{+0.45}_{-0.43}\pm0.22\pm0.23\pm0.60)\times10^{-7}.
\end{equation}

Hence, considering the status of theoretical calculation and the recent improvement in experimental measurement, we believe that it is valuable to re-consider the decays of $\bar{B}_s^0\to \phi l^+ l^-$ ($l=e, \mu, \tau$) theoretically. In this work, we revisit the form factors in $\bar{B}_s^0\to \phi$ transition in QCD sum rule, and use these form factors to calculate the branching ratios of $\bar{B}_s^0\to \phi e^+ e^-$, $\bar{B}_s^0\to \phi \mu^+ \mu^-$ and $\bar{B}_s^0\to \phi \tau^+ \tau^-$. Finally, we compare our results of form factors and branching ratios with previous theoretical works as well as the latest experimental data.

The paper is organized as followings. In Sec. II, we present the effective Hamiltonian and effective amplitude
of $\bar{B}_s^0\to \phi l^+ l^-$ decay. Section III $\thicksim$ IV are devoted to the calculation of the form factors in QCD sum rule method. Section V is for the numerical analysis and discussion. Finally, a brief summary is presented in Sec. VI.


\section{Effective Hamiltonian}
At quark level, the effective Hamiltonian of the rare semileptonic decay $b \to s l^+ l^-$ can be written as \cite{hamiltonian},
\begin{equation}\label{2}
\mathcal{H}_{{\rm eff}}=-\frac{G_F}{\sqrt{2}}V_{tb}V_{ts}^*\sum_{i=1}^{10}{C_i(\mu)O_i(\mu)},
\end{equation}
where $V_{tb}V_{ts}^*$ is the product of relevant CKM matrix elements. $C_i$ denotes Wilson coefficient, and the operators $O_i$ are
\begin{align*}
  &Q_1=(\bar{s}_\alpha c_\beta)_{V-A}(\bar{c}_\beta b_\alpha)_{V-A},
  &Q_2&=(\bar{s}c)_{V-A}(\bar{c}b)_{V-A},\\
  &Q_3=(\bar{s}b)_{V-A}\sum_q (\bar{q}q)_{V-A},
  &Q_4&=(\bar{s}_\alpha b_\beta)_{V-A}\sum_q (\bar{q}_\beta q_\alpha)_{V-A},\\
  &Q_5=(\bar{s}b)_{V-A}\sum_q (\bar{q}q)_{V+A},
  &Q_6&=(\bar{s}_\alpha b_\beta)_{V-A}\sum_q (\bar{q}_\beta q_\alpha)_{V+A},\\
  &Q_7=\frac{\alpha_{e}}{2\pi}m_b\bar{s}_\alpha \sigma^{\mu\nu}(1+\gamma^5)b_\alpha F_{\mu\nu},
  &Q_8&=\frac{\alpha_s}{2\pi}m_b\bar{s}_\alpha \sigma^{\mu\nu}(1+\gamma^5)T_{\alpha\beta}^a
    b_\beta G_{\mu\nu}^a,\\
  &Q_9=\frac{\alpha}{2\pi}(\bar{s}b)_{V-A}(\bar{l}l)_V,
  &~Q_{10}&=\frac{\alpha}{2\pi}(\bar{s}b)_{V-A}(\bar{l}l)_A.\\
\end{align*}
Then the effective Hamiltonian above leads to the following decay amplitude of $\bar{B}_s^0\to \phi l^+ l^-$ \cite{hamiltonian}
\begin{eqnarray}\label{3}
\mathcal{M}(\bar{B}_s^0\to \phi l^+ l^-)&=&\frac{G_F \alpha}{2\sqrt{2}\pi}
V_{tb}V_{ts}^*\left[C^{{\rm eff}}_9\langle\phi(\varepsilon,p_2)|\bar{s}\gamma_\nu (1-\gamma_5)b|\bar{B}_s^0(p_1)
\rangle\bar{\ell}\gamma^\nu\ell\right.  \nonumber \\
&~&\left.+C_{10}\langle\phi(\varepsilon,p_2)|\bar{s}\gamma_\nu (1-\gamma_5)b|\bar{B}_s^0(p_1)\rangle\bar{\ell}\gamma^\nu\gamma_5\ell\right. \\
&~&\left.-2C^{{\rm eff}}_7 m_b \frac{{\rm i}}{q^2}\langle\phi(\varepsilon,p_2)|\bar{s} \sigma_{\nu\lambda}q^\lambda(1+\gamma_5)b|\bar{B}_s^0(p_1)\rangle\bar{\ell}\gamma^\nu\ell\right] \nonumber
\end{eqnarray}
where $p_1$ and $p_2$ are momenta of $\bar{B}_s^0$ and $\phi$ mesons, respectively.
$q$ is the momentum transfer $q=p_1-p_2$. $C^{{\rm eff}}_9$ and $C^{{\rm eff}}_7$ are two effective Wilson coefficients, with $C^{{\rm eff}}_7=C_7-{C_5}/3-C_6$. As for the effective Wilson coefficient $C^{{\rm eff}}_9$, we take the expression in Ref. \cite{hamiltonian},
which is given as followings
\begin{eqnarray} \label{c9eff}
C^{{\rm eff}}_9&=&C_9+C_0\left[h(\hat{m}_c,\hat{s})+\frac{3\pi\kappa}{\alpha^2}\sum_{V_i=\psi(1s;2s)}\frac{\Gamma(V_i\to l^+l^-)m_{V_i}}{m^2_{V_i}-q^2-{\rm i}m_{V_i}\Gamma_{V_i}}\right]\nonumber \\
&-&\frac{1}{2}h(1,\hat{s})(4C_3+4C_4+3C_5+C_6)\\
&-&\frac{1}{2}h(0,\hat{s})(C_3+3C_4)+\frac{2}{9}(3C_3+C_4+3C_5+C_6),       \nonumber
\end{eqnarray}
where we define
$$C_0=3C_1+C_2+3C_3+C_4+3C_5+C_6,$$
$$h(0,\hat{s})=\frac{8}{27}-\frac{8}{9}\ln{\frac{m_b}{\mu}}-\frac{4}{9}\ln{\hat{s}}+{\rm i}\pi\frac{4}{9}, $$
and
$$
h(\hat{m}_c,\hat{s})=-\frac{8}{9}\ln{\frac{m_b}{\mu}}-\frac{8}{9}\ln{\hat{m}_c}+\frac{8}{27}+\frac{4}{9}x-\frac{2}{9}(2+x)|1-x|^{\frac{1}{2}}
\left\{
\begin{array}{rcl}
&&(\ln{|\frac{\sqrt{1-x}+1}{\sqrt{1-x}-1}|-{\rm i}\pi}),  ~{x < 1}\\
&&2\arctan{\frac{1}{\sqrt{x-1}}}~~~~~, ~{x > 1}
\end{array} \right.
,$$
with $x={4{\hat{m}}_c}^2/{\hat{s}}$,
$\hat{m}_c={m_c}/{m_{B_s}}$, $\hat{s}={q^2}/{m^2_{B_s}}$, $\kappa={1}/{C_0}$ and $\mu=m_b$.

\section{Form Factors from QCD Sum Rule}
We have calculated the hadronic matrix elements
$\langle\phi(\varepsilon,p_2)|\bar{s}\gamma_\nu (1-\gamma_5)b|\bar{B}_s^0(p_1)\rangle$ in the
decay amplitude given in Eq.~(\ref{3}) in our previous work \cite{PYQ}. So in this work, we need only to deal with
the other hadronic matrix element $\langle\phi(\varepsilon,p_2)|\bar{s}\sigma_{\nu\lambda} q^\lambda (1+\gamma_5)b|\bar{B}_s^0(p_1)\rangle$
in Eq.~(\ref{3}).

Similarly the hadronic matrix element $\langle \phi|\bar{s}\sigma_{\nu\lambda} q^\lambda (1+\gamma_5)b|\bar{B}_s^0 \rangle$ can be decomposed as \cite{Lu/form-factors}
\begin{eqnarray}\label{5}
\langle\phi(\varepsilon,p_2)|\bar{s}\sigma_{\nu\lambda} q^\lambda (1+\gamma_5)b|\bar{B}_s^0(p_1)\rangle
&=&2{\rm i}\varepsilon_{\nu\rho\alpha\beta}\varepsilon^{*\rho}p_1^\alpha
p_2^\beta  T_1(q^2) \nonumber \\
&&+[\varepsilon^*_\nu(m_{B_s}^2-m_{\phi}^2)-(\varepsilon^*\cdot q)(p_1+p_2)_\nu] T_2(q^2) \\
&&+(\varepsilon^*\cdot q)[q_\nu-\frac{q^2}{m_{B_s}^2-m_{\phi}^2}(p_1+p_2)_\nu]
 T_3(q^2),\nonumber
\end{eqnarray}
where $T_1$, $T_2$ and $T_3$ are the transition form factors associated with
the current of $j^T_\nu (0)=\bar{s}\sigma_{\nu\lambda} q^\lambda (1+\gamma_5)b$.

As what we did in Ref. \cite{PYQ}, at first we consider a three-point correlation function that is defined as
\begin{equation}\label{6}
\Pi_{\mu\nu}={\rm i}^2\int {\rm d}^4 x {\rm d}^4 y {\rm e}^{{\rm i}p_2\cdot x-{\rm i}p_1\cdot y}
 \langle 0|T\{j^\phi_\mu(x) j^T_\nu (0)j_5(y)\} |0\rangle ,
\end{equation}
where $j_\mu^\phi(x)=\bar{s}(x)\gamma_\mu s(x)$, $j^T_\nu (0)=\bar{s}\sigma_{\nu\lambda} q^\lambda (1+\gamma_5)b$ and
$j_5(y)=\bar{b}(y){\rm i}\gamma_5 s(y)$, which are the current of $\phi$ channel,
the current of weak transition and the current of $\bar{B}_s^0$ channel, respectively.

Next we reexpress the correlation function by using the double dispersion relation
\begin{equation}\label{7}
\Pi_{\mu\nu}=\int {\rm d}s_1 {\rm d}s_2\frac{\rho (s_1,s_2,q^2)}{(s_1-p_1^2) (s_2-p_2^2)},
\end{equation}
where the spectral density function $\rho (s_1,s_2,q^2)$ can be expressed as the form containing
a full set of intermediate hadronic states as shown below,
\begin{equation}\label{rho}
\rho (s_1,s_2,q^2)=\sum_{X}\sum_{Y}\langle 0|j_\mu^\phi |X\rangle \langle
   X|j^T_\nu |Y\rangle \langle Y|j_5 |0 \rangle \delta (s_1-m_Y^2)
    \delta (s_2-m_X^2)\theta(p_X^0)\theta(p_Y^0),
\end{equation}
where $X$ and $Y$ denote the full set of hadronic states of
$\phi$ and $\bar{B}^0_s$ channels, respectively. According to Eqs. (\ref{7}) and (\ref{rho}), we can integrate
over $s_1$ and $s_2$, then separate the ground states, excited states and continuum states, the correlation
function can be expressed as
\begin{eqnarray}\label{9}
\Pi_{\mu\nu}=\frac{ m_\phi f_\phi
  \varepsilon_\mu^{(\lambda)}\langle
   \phi(\varepsilon_\mu^{(\lambda)},p_2)|j^T_\nu |\bar{B}_s^0(p_1)\rangle
   f_{B_s} m_{B_s}^2}{(m_{B_s}^2-p_1^2)(m_{\phi}^2-p_2^2)(m_b+m_s)}\nonumber\\
    +   \mbox{\small{excited and continuum states}}.
    \label{16}
\end{eqnarray}
In the above equation, we have used the following definition of relevant matrix elements
 \begin{eqnarray}\label{10}
&&\langle 0|\bar{s}\gamma_\mu s|\phi\rangle =m_\phi f_\phi
  \varepsilon_\mu^{(\lambda)}, \nonumber \\
&&\langle 0|\bar{s}{\rm i}\gamma_5 b|\bar{B}_s^0\rangle =
\frac{f_{B_s}m_{B_s}^2}{m_b+m_s},
\end{eqnarray}
where $f_\phi$ and $f_{B_s}$ are decay constants of the relevant mesons. In principle,
$\phi$ and $\omega$ can mix via strong interaction, the mixing angle $\delta$ between nonstrange and
strange quark wave function has been analyzed to be $\delta=-(3.34\pm 0.17)^\circ$ \cite{Benayoun1,Kucu,Gronau1,Benayoun2,Gronau2}, which shows that $\phi$ meson is
dominated by component $s\bar{s}$. Therefore, we can safely drop the mixing effect of $\omega-\phi$ in $D_s\to\phi$
transition process, and $\phi$ meson is treated as $s\bar{s}$ component, which is referred to as ideal mixing.

By taking the operator product expansion (OPE) for the time-ordered current operator
in Eq. (\ref{6}), we can get another expression for the correlation function in
terms of Wilson coefficients and condensates of local operators
\begin{eqnarray}\label{11}
\Pi_{\mu\nu}&=&{\rm i}^2\int {\rm d}^4x {\rm d}^4 y {\rm e}^{{\rm i}p_2\cdot x-{\rm i}p_1\cdot y}
 \langle 0|T\{j^\phi_\mu(x) j^T_\nu (0)j_5(y)\}|0\rangle \nonumber\\
 &=&C_{0\mu\nu} I +C_{3\mu\nu} \langle 0|\bar{\Psi}\Psi|0\rangle
    +C_{4\mu\nu} \langle 0|G^a_{\alpha\beta}G^{a\alpha\beta}|0\rangle
    +C_{5\mu\nu} \langle 0|\bar{\Psi}\sigma_{\alpha\beta}T^a G^{a\alpha\beta}\Psi|0\rangle
    \nonumber\\
  &~+&C_{6\mu\nu}\langle 0|
 \bar{\Psi}\Gamma \Psi \bar{\Psi}\Gamma^{\prime}\Psi|0\rangle +\cdots,
 \label{18}
\end{eqnarray}
where $C_{i\mu\nu}$ denotes Wilson coefficients. $I$, $\bar{\Psi}\Psi$ and $G^a_{\alpha\beta}$ are the unit
operator, the local fermion field operator of light quarks and the gluon strength tensor, respectively.
$\Gamma$ and $\Gamma^{\prime}$ are the matrices that appear in the calculation of Wilson's coefficients.
From the Lorentz structure of the correlation function, we can know that Eq.~(\ref{11}) can be rewritten as
\begin{equation}\label{12}
\Pi_{\mu\nu}={\rm i}\kappa_0\varepsilon_{\mu\nu\alpha\beta}p_1^\alpha
p_2^\beta+(\kappa_1 p_{1\mu}p_{1\nu}+\kappa_2 p_{2\mu}p_{2\nu}+\kappa_3
p_{1\mu}p_{2\nu}+\kappa_4 p_{1\nu}p_{2\mu}+\kappa_5g_{\mu\nu}).
\end{equation}
The coefficients $\kappa_i$'s contain perturbative and condensate contributions
\begin{equation}\label{13}
\kappa_i=\kappa_i^{{\rm pert}}+\kappa_i^{(3)}+\kappa_i^{(4)}+\kappa_i^{(5)}+\kappa_i^{(6)}+\cdots,
\end{equation}
where $\kappa_i^{{\rm pert}}$ is the perturbative contribution, and $\kappa_i^{(3)}$, $\kappa_i^{(4)}$, $\kappa_i^{(5)}$, $\kappa_i^{(6)}$, $\cdots$ are
contributions of condensates of operators with increasing dimension in OPE.

Since the  perturbative contribution and gluon-condensate contribution contain the loop integral of momentum,
we can obtain the dispersion integrals of $\kappa_i^{{\rm pert}}$ and $\kappa_i^{(4)}$, which can be expressed as
\begin{eqnarray} \label{dss}
 \kappa_i^{{\rm pert}}&=&\int^{\infty}_{s_1^L} {\rm d} s_1
 \int^{\infty}_{s_2^L} {\rm d} s_2\frac{\rho^{{\rm pert}}_i(s_1,s_2,q^2)}{(s_1-p_1^2)(s_2-p_2^2)},
 \nonumber  \\
\kappa_i^{(4)}&=&\int^{\infty}_{s_1^L} {\rm d} s_1
 \int^{\infty}_{s_2^L}{\rm d} s_2\frac{\rho^{(4)}_i(s_1,s_2,q^2)}{(s_1-p_1^2)(s_2-p_2^2)},
  \end{eqnarray}
where $s_1^L$ and $s_2^L$ are the lower limits of $s_1$ and $s_2$, respectively, which can be found in Appendix A.
In principle Eqs.~(\ref{9}) and (\ref{12}) should be equivalent to each other, because they are two different
expressions for the same correlation function $\Pi_{\mu\nu}$. By using the assumption of quark-hadron duality
\cite{SVZ1,SVZ2}, one can approximate the contribution of the higher excited and continuum states in $\Pi_{\mu\nu}$
in Eq. (\ref{9}) as the integration of $\int ds_1 ds_2$ in Eq. (\ref{dss}) over some thresholds $s_1^0$ and $s_2^0$. Then one can get rid of the contribution of the higher excited and continuum states in Eq.~(\ref{9}),
and obtain an equation for the form factors by equating Eqs.~(\ref{9}) and (\ref{12}),
where Eq. (\ref{dss}) should be replaced as
\begin{eqnarray}
 \kappa_i^{{\rm pert}}&=&\int^{s_1^0}_{s_1^L} {\rm d} s_1
 \int^{s_2^0}_{s_2^L} {\rm d} s_2\frac{\rho^{{\rm pert}}_i(s_1,s_2,q^2)}{(s_1-p_1^2)(s_2-p_2^2)},
 \nonumber  \\
\kappa_i^{(4)}&=&\int^{s_1^0}_{s_1^L} {\rm d} s_1
 \int^{s_2^0}_{s_2^L}{\rm d} s_2\frac{\rho^{(4)}_i(s_1,s_2,q^2)}{(s_1-p_1^2)(s_2-p_2^2)}.
 \end{eqnarray}
In order to improve the equation, Borel transformation needs to be introduced, that is, for any function $f(x^2)$,
$$\hat{B}_{\left|\frac{}{}\right.x^2,M^2}f(x^2)=\lim_{\small\begin{array}{cc}& k\to\infty, x^2\to -\infty
\\&-x^2/k= M^2  \end{array} } \frac{(-x^2)^k}{(k-1)!}\frac{\partial ^k}{\partial (x^2)^k}
    f(x^2).$$
Borel transformation can suppress both the contribution of higher excited states and contributions of
operators of higher dimension in OPE. Then matching these two forms of the correlation
function in Eqs.~(\ref{9}) and (\ref{12}), and performing
Borel transformation for both variables $p_1^2$ and $p_2^2$, QCD sum rules for these three form factors
related to matrix hadronic element $\langle\phi(\varepsilon,p_2)|\bar{s}\sigma_{\nu\lambda} q^\lambda (1+\gamma_5)b|\bar{B}_s^0(p_1)\rangle$
can be obtained
\begin{eqnarray}\label{14}
T_1(q^2)&=& \frac{(m_b+m_s)}{2m_\phi f_\phi
 f_{B_s}m_{B_s}^2}{\rm e}^{m_{B_s}^2/M_1^2}{\rm e}^{m_{\phi}^2/M_2^2}M_1^2M_2^2
 \cdot \hat{B}\kappa_0, \nonumber \\
T_2(q^2)&=& -\frac{(m_b+m_s)}{m_\phi f_\phi
 f_{B_s}m_{B_s}^2(m_{B_s}^2-m_{\phi}^2)}{\rm e}^{m_{B_s}^2/M_1^2}{\rm e}^{m_{\phi}^2/M_2^2}M_1^2M_2^2
 \cdot \hat{B}\kappa_5, \\
T_3(q^2)&=& -\frac{(m_b+m_s)}{m_\phi f_\phi
 f_{B_s}m_{B_s}^2}{\rm e}^{m_{B_s}^2/M_1^2}{\rm e}^{m_{\phi}^2/M_2^2}M_1^2M_2^2
 \cdot \frac{1}{2}\hat{B}(\kappa_1-\kappa_3), \label{21}  \nonumber
 \end{eqnarray}
where $\hat{B} \kappa_i$ denotes Borel transformation of $\kappa_i$ for both
variables $p_1^2$ and $p_2^2$. $M_1$ and $M_2$ are Borel
parameters.

\section{ The Calculation of the Wilson Coefficients}
\begin{figure}[h]
\begin{center}
\begin{picture}(300,200)(-15,-25.98)
\DashLine(-11.25,100.515)(0,120){3} \Line(0,120)(30,171.96)
\Line(0,120)(30,171.96)\Line(30,171.96)(60,120)\Line(60,120)(0,120)
\DashLine(60,120)(71.25,100.515){3}\Photon(30,171.96)(30,190.71){1}{4}

\GlueArc(0,120)(30,40,60){2}{2}\GlueArc(0,120)(30,0,20){2}{2}
\put(19.88,137,25){$\times$}\put(23.5,129){$\times$}

\DashLine(88.75,100.515)(100,120){3} \Line(100,120)(130,171.96)
\Line(100,120)(130,171.96)\Line(130,171.96)(160,120)\Line(160,120)(100,120)
\DashLine(160,120)(171.25,100.515){3}\Photon(130,171.96)(130,190.71){1}{4}

\GlueArc(130,171.96)(30,240,260){2}{2}\GlueArc(130,171.96)(30,280,300){2}{2}
\put(120.4,138.8){$\times$}\put(131.3,138.8){$\times$}

\DashLine(188.75,100.515)(200,120){3} \Line(200,120)(230,171.96)
\Line(200,120)(230,171.96)\Line(230,171.96)(260,120)\Line(260,120)(200,120)
\DashLine(260,120)(271.25,100.515){3}\Photon(230,171.96)(230,190.71){1}{4}

\GlueArc(260,120)(30,160,180){2}{2}\GlueArc(260,120)(30,120,140){2}{2}
\put(232.3,136){$\times$}\put(228.125,128){$\times$}

\DashLine(-11.25,-19.485)(0,0){3} \Line(0,0)(30,51.96)
\Line(0,0)(30,51.96)\Line(30,51.96)(60,0)\Line(60,0)(0,0)
\DashLine(60,0)(71.25,-19.485){3}\Photon(30,51.96)(30,70.71){1}{4}

\Gluon(11.25,19.485)(-1.74,26.985){2}{2}\Gluon(18.75,32.475)(5.76,39.975){2}{2}

\put(-5.625,23.5){$\times$}\put(1.65,36.5){$\times$}
 \DashLine(88.75,-19.485)(100,0){3} \Line(100,0)(130,51.96)
\Line(100,0)(130,51.96)\Line(130,51.96)(160,0)\Line(160,0)(100,0)
\DashLine(160,0)(171.25,-19.485){3}\Photon(130,51.96)(130,70.71){1}{4}

\Gluon(122.5,0)(122.5,-15){2}{2}\Gluon(137.5,0)(137.5,-15){2}{2}

\put(119.875,-16.5){$\times$}\put(135.25,-16.5){$\times$}

\DashLine(188.75,-19.485)(200,0){3} \Line(200,0)(230,51.96)
\Line(200,0)(230,51.96)\Line(230,51.96)(260,0)\Line(260,0)(200,0)
\DashLine(260,0)(271.25,-19.485){3}\Photon(230,51.96)(230,70.71){1}{4}

\Gluon(248.75,19.485)(261.74,26.985){2}{2}\Gluon(241.25,32.475)(254.24,39.975){2}{2}

\put(256.5,24.75){$\times$}\put(250,37.5){$\times$}

\put(23,90){(a)}
 \put(125,90){(b)}
 \put(225,90){(c)}

 \put(23,-35){(d)}
 \put(125,-35){(e)}
 \put(225,-35){(f)}
\end{picture}
\end{center}
\vspace{0.5cm}
\caption{\label{gg}\small Diagrams for contributions of gluon-gluon operator.}
\end{figure}
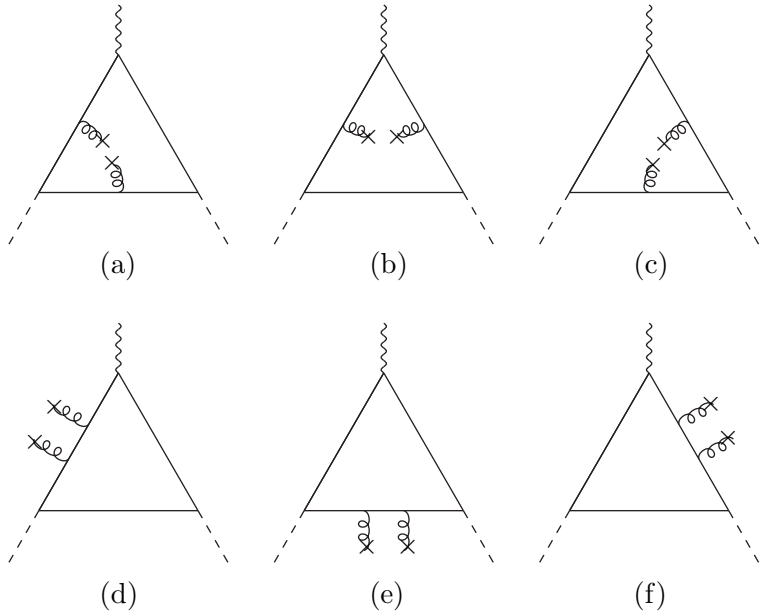

In this section, we discuss the calculation of Wilson Coefficients in the OPE. The diagrams to be considered here
are similar to that used in our previous work in Ref. \cite{PYQ}.
The difference is that the weak transition current $j_\nu (0)=\bar{s}\gamma_\nu (1-\gamma_5)b$
is replaced by the tensor current $j^T_\nu (0)=\bar{s}\sigma_{\nu\lambda} q^\lambda (1+\gamma_5)b$
appearing in Eq.~(\ref{3}).

Here we only depict the diagrams for contributions of gluon-gluon operator in Fig.\ref{gg},
because our calculation shows that the contribution of gluon-gluon operator does not
completely cancel out for the tensor current, which is different from the case of $V-A$ current.
But the contributions of these diagrams are very small compared with other operators.
Different from the treatment in Ref. \cite{R.K-SumR}, we do not ignore these contributions
in the following calculations.

The cancellation of the contribution of gluon-gluon operator for the case of
$V-A$ current seems not because of any symmetry principle. It is only only because, in the
fixed-point gauge the color field can be expanded as $A^a_\mu (z)=\int^1_0d\beta \beta z^\rho
G^a_{\rho\mu}(\beta z)=\frac{1}{2}z^\rho  G^a_{\rho\mu}(0)+\cdots$
at leading order, only at leading order the contribution of gluon-gluon operator vanish.
If the higher order in the expansion
$ A^a_{\mu}(z)=\frac{1}{2}z^\rho G^a_{\rho\mu}(0)+\frac{1}{3}z^\alpha z^\rho
\hat{D}_\alpha G^a_{\rho\mu}(0)+\cdots ~$
is considered, the contribution may not vanish for the case of V-A current,
but it must be small because of the short-distance nature of Wilson coefficients.

The final results of Borel transformed coefficients
$\hat{B}\kappa_0$, $\hat{B}(\kappa_1-\kappa_3)$ and $\hat{B}\kappa_5$ in
Eq. (\ref{14}) are given in Appendix A.
\section{Numerical analysis and discussion}
The input parameters required for numerical calculation are taken as followings \cite{SVZ1,SVZ2,Colan2000}:
\begin{eqnarray}\label{38}
&\langle \bar{q}q\rangle =-(0.24\pm 0.01 \mbox{GeV})^3, ~~~~
\langle \bar{s}s\rangle =(0.8\pm 0.2)\langle \bar{q}q\rangle,
\nonumber\\[4mm]
 & g\langle \bar{\Psi}\sigma TG\Psi \rangle =m_0^2 \langle
\bar{\Psi}\Psi \rangle, ~~~~\alpha_s\langle \bar{\Psi}\Psi\rangle
^2= 6.0\times10^{-5}\mbox{GeV}^6 ,\\[4mm]
& \alpha_s\langle GG\rangle
= 0.038\mbox{GeV}^4, ~~~~m_0^2=0.8\pm 0.2 \mbox{GeV}^2. \nonumber
\end{eqnarray}

The standard values of the condensates above at the renormalization point $\mu =1\mbox{GeV}$ are
from Refs. \cite{SVZ1,SVZ2,Colan2000}, and the relevant mass parameters and decay constants are \cite{PDG2018,fBs},
\begin{eqnarray}
&&m_s=95\mbox{MeV},~~~~~~~~~~m_b=4.18\mbox{GeV},~~~~~~~~~~m_e=0.511\mbox{MeV}, \nonumber \\
&&m_{\mu}=0.106\mbox{GeV},~~~~~~~m_{\tau}=1.777\mbox{GeV},~~~~~~~~m_\phi=1.02\mbox{GeV}, \nonumber\\
&&m_{B_s}=5.367\mbox{GeV},~~~~~m_{J/\psi}=3.097\mbox{GeV},~~~~~~m_{\psi^{'}}=3.686\mbox{GeV},\nonumber \\
&&f_{B_s}=0.266\pm0.019\mbox{GeV},~~~~~~~f_{\phi}=0.228\mbox{GeV}.~~~~~~~~~~~~~~~~~~~~~~~~~~
\end{eqnarray}
Other parameters to be used include \cite{PDG2018}:
\begin{eqnarray}
G_F=1.1663787\times10^{-5}\mbox{GeV}^{-2}, ~~~\alpha=7.297\times10^{-3},
~~~|V^*_{ts}V_{tb}|=0.039741,
\end{eqnarray}
and the threshold parameters $s_1^0$ and $s_2^0$ for $\bar{B}^0_s$ and $\phi$ mesons are
\begin{equation}
s_1^0=34.9\thicksim35.9\mbox{GeV}^2, ~~~~s_2^0=1.9\thicksim2.1\mbox{GeV}^2.
\end{equation}

For the Wilson coefficients appearing in Eq.(\ref{c9eff}) that are
involved in our numerical calculation, the values are listed in Table \ref{Table:1} \cite{B/Vll,Wilson-Co}.
\begin{table}[h]
\begin{center}
\caption{\label{Table:1}Wilson coefficients (at renormalization scale $\mu=m_b$)}
\begin{tabular}{|c|c|c|c|c|c|c|c|c|}\hline
$C_1$&
$C_2$ & $C_3$ & $C_4$& $C_5$& $C_6$ &$C^{eff}_7$&
$C_9$ & $C_{10}$\\ \hline $~-0.176~$ &$~1.078~$ & $~0.014~$ & $~-0.034~$ & $~0.008~$& $~-0.039~$ &$~-0.313~$ &$~4.344~$ & $~-4.669~$\\ \hline
\end{tabular}\end{center}
\end{table}

\begin{figure}[tbp]
  \centering
  \includegraphics[width=0.8\textwidth,origin=l,angle=0]{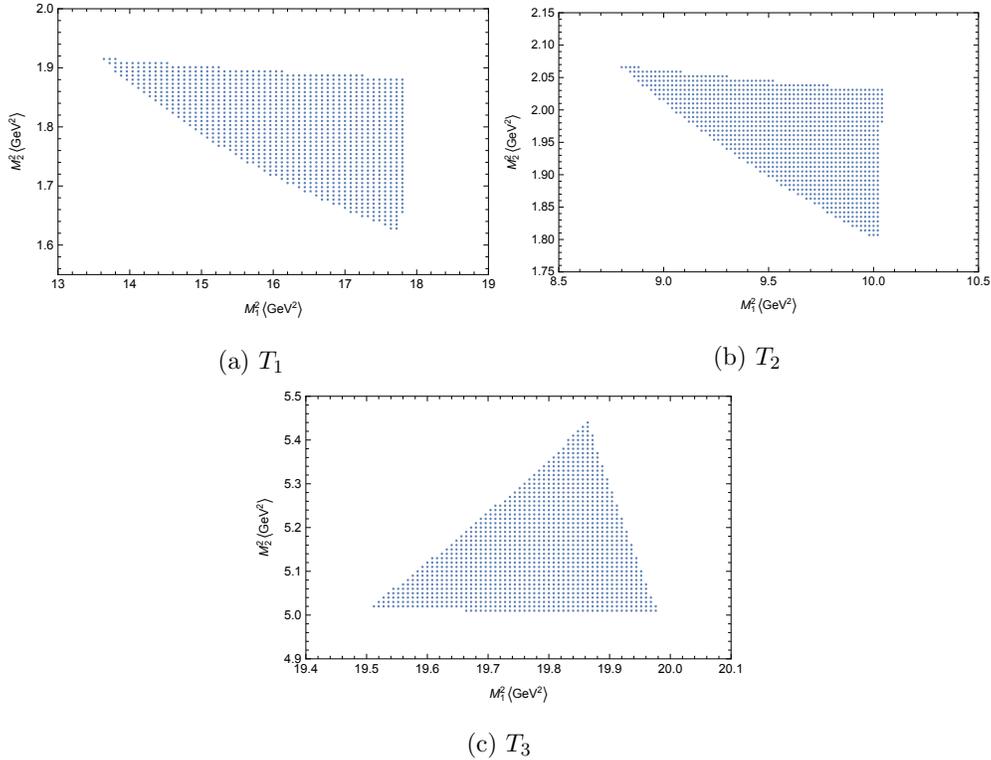}
  \caption{Selected stability regions of $M_1^2$ and $M_2^2$.} \label{fig2}
\end{figure}

Next we need to select the appropriate regions for Borel parameters $M_1$ and $M_2$.
In our previous works \cite{PYQ,dly,yang}, we have discussed the selection of Borel parameters in detail.
So we do not repeat the details in this paper. The requirements to select Borel Parameters are
directly given in Table \ref{Ta2}, and the selected two-dimensional region for $M_1$ and $M_2$ are depicted in Fig.\ref{fig2}.
\begin{table}[h]
\caption{Requirements to select Borel Parameters $M_1^2$ and $M_2^2$
 for each form factors $T_1(0)$, $T_2(0)$ and $T_3(0)$}
\begin{center}
\begin{tabular}{|c|c|c|c|}\hline
Form Factors & contribution  & continuum of & continuum of  \\
             &of condensate &  $\bar{B}^0_s$ channel & $\phi$ channel\\
             \hline
$T_1(0)$&
$\le 54.4\% $ & $\le 15.5\%$ & $\le 56\%$ \\ \hline $T_2(0)$ &$\le
54.4\% $ & $\le 12\%$ & $\le 56\%$ \\ \hline $T_3(0)$  & $\le 55.4\% $
& $\le 41.2\%$ & $\le 56.8\%$ \\ \hline
\end{tabular}\end{center}
\label{Ta2}
\end{table}

After numerical analysis, the final results for the form factors at $q^2=0$ are
\begin{eqnarray}\label{18f}
&&T_1(0)~=~0.33\pm 0.07,\nonumber
\\
&&T_2(0)~=~0.33\pm 0.07,
\\
&&T_3(0)~=~0.22\pm 0.05,\nonumber
\end{eqnarray}
where the errors are estimated by the uncertainty of
the standard values of the condensates, the
variation of the threshold parameters $s^0_1$ and $s^0_2$,
the variation of Borel parameters, and the
variation of the other input parameters. The error caused by the
uncertainty of the condensates is about 25$\%$ of the central value of the form factors, the
error caused by the variation of the threshold parameters $s^0_{1,2}$ is about 5$\%$ of the central value,
the error caused by the variation of Borel parameters is about 6$\%$ of the central value, and
the error caused by the uncertainty of the other input parameters is less than a few percent.
All the errors are added quadratically. In addition,
the $b$ quark mass given by Ref. \cite{PDG2018} is $m_b=4.18^{+0.04}_{-0.03}$.
The error caused by the uncertainty of $b$ quark mass is about 0.8$\%$, which is much
smaller than the errors caused by the other sources.

The comparison of the form factors obtained in this work in Eq.(\ref{18f}) with other theoretical results
calculated by LCSR in Ref. \cite{LCSR}, CQM in Ref. \cite{CQM-Me}, RQM in Ref. \cite{FG}, and also in QCD sum rule in
Ref. \cite{R.K-SumR} are shown in Table \ref{Ta3}. Some of the form factors obtained in
Ref. \cite{R.K-SumR} are different from others by a negative sign. This will affect the physical
results of the differential decay width of $\bar{B}_s^0\to \phi l^+ l^-$.
By comparison, we find that the results of $T_1(0)$, $T_2(0)$ and $T_3(0)$ in our work, especially the value of $T_3(0)$,
are more consistent with the results obtained by LCSR method in Ref. \cite{LCSR} within the range of uncertainty.
Comparing the OPE coefficients in Ref. \cite{R.K-SumR} with the relevant coefficients in this work, we find
that the reason for the difference is that there is
no contribution of ${m_b}/{M^2_1 M^2_2}$ and ${m_s}/{M^2_1 M^2_2}$ in Ref. \cite{R.K-SumR}.
The contribution of these two types of terms comes from the first term in the right side of Eq. (\ref{q-q}) \cite{Colan2000,dly},
which gives the main contribution in our calculation
\begin{eqnarray}
&&\langle 0 |\bar{\Psi}^a_\alpha (x)\Psi^b_\beta (y)|0\rangle
 =\delta_{ab}\left[ \langle \bar{\Psi} \Psi\rangle\left(
   \frac{1}{12}\delta_{\beta\alpha}+{\rm i}\frac{m}{48}
   (\not{x}-\not{y})_{\beta\alpha}-\frac{m^2}{96}
 (x-y)^2\delta_{\beta\alpha}\right.\right. \nonumber\\
   &&\left.\left.-\frac{{\rm i}}{3!}\frac{m^3}{96}
 (x-y)^2(\not{x}-\not{y})_{\beta\alpha} \right)
 +g\langle \bar{\Psi}\sigma TG\Psi\rangle\left(\frac{1}{192}(x-y)^2
 \delta_{\beta\alpha}\right.\right. \nonumber\\
   &&\left.\left.+\frac{{\rm i}}{3!}\frac{m}{192}
 (x-y)^2(\not{x}-\not{y})_{\beta\alpha}\right)
 -\frac{{\rm i}}{3!}\frac{g^2}{3^4\times 2^4}\langle
 \bar{\Psi}\Psi\rangle ^2 (x-y)^2(\not{x}-\not{y})_{\beta\alpha}
 \right. \nonumber \\&& \left.\frac{}{}+\cdots \right].
 \label{q-q}
 \end{eqnarray}
Moreover, the contribution of the operator of dimension-5 is greater than
that of the operator of dimension-3 in Ref. \cite{R.K-SumR}, which is also
different from our calculation.

\begin{table}[h]
\caption{Comparison of our results of form factors with other works}
\begin{center}
\begin{tabular}{|c|c|c|c|}
\hline
 &$T_1(0)$ & $T_2(0)$ &$T_3(0)$  \\
                         \hline
$\mbox{LCSR}$ \cite{LCSR}  & $0.35 $ & $0.35$ & $0.18$
\\ \hline
$\mbox{CQM}$  \cite{CQM-Me} & $0.38 $&$0.38$ & $0.26$  \\ \hline
$\mbox{RQM}$  \cite{FG} & $0.275 $ &$0.275$ & $0.133$  \\ \hline
$\mbox{SR}$ \cite{R.K-SumR} & $-0.35$ & $0.37$ & $-0.28$  \\ \hline
$\mbox{This work}$    & $~~0.33\pm0.07~~ $ & $~~0.33\pm0.07~~$ &
$~~0.22\pm0.05~~$  \\ \hline
\end{tabular}\end{center}
\label{Ta3}
\end{table}

\begin{figure}[tbp]
  \centering
  \includegraphics[width=0.5\textwidth,origin=l,angle=0]{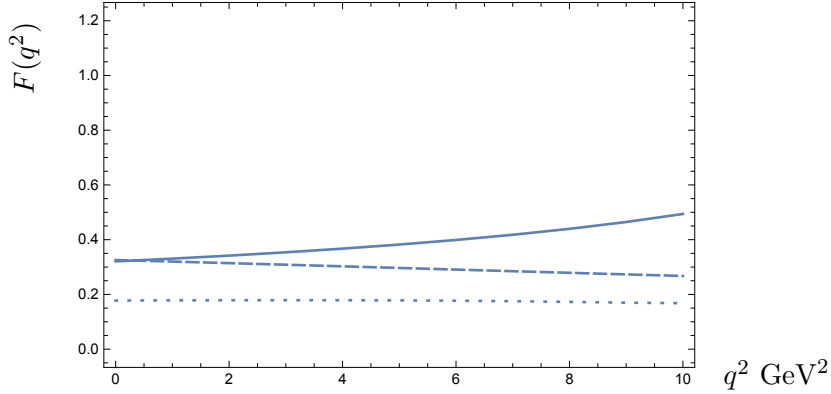}
  \put(10,0){\small{$q^2~\mbox{GeV}^2$}}
  \put(-260,110){\rotatebox{90}{\small{$F(q^2)$} }}
  \caption{\small $q^2$-dependence of the form factors from QCD sum rule.
The solid curve is for $T_1(q^2)$, the dashed curve for $T_2(q^2)$,
and the dotted curve for $T_3(q^2)$.} \label{q02}
\end{figure}

The physical region for $q^2$ in $\bar{B}_s^0\to \phi l^+l^-$ decay is: $(2m_l)^2 \leq q^2 \leq (m_{B_s}-m_\phi)^2$.
The $q^2$-dependence of the form factors within this range is shown in Fig. \ref{q02} using the central
values of the input parameters. We can find that the $q^2$-dependence of $T_1(q^2)$ calculated in QCD sum rule can be well fitted by the single-pole model
\begin{equation}
T_1(q^2)=\frac{T_1(0)}{1-q^2/(m_{{\rm pole}}^{T_1})^2},
\end{equation}
while the $q^2$-dependences of $T_2(q^2)$ and $T_3(q^2)$ are very weak, so we can take $T_2(q^2)=T_2(0)$, $T_3(q^2)=T_3(0)$ as approximations. The weak dependence of $T_2(q^2)$ and $T_3(q^2)$ on $q^2$ stems from the mutual cancellation of the
perturbative contribution and the condensate contribution. For $T_2(q^2)$, the perturbative contribution increases as $q^2$
being large, while the contribution of condensates decreases, and as a sum the $q^2$-dependence cancel mostly. For $T_3(q^2)$, the perurbative contribution decreases while the condensates contribution increases as $q^2$ being large.
This is similar to the behavior of the form factors for $D$ decays found in Ref. \cite{BBD}. The weak dependence of
$T_{2,3}(q^2)$ on $q^2$ calculated from QCD sum rule implies that the assumption of single-pole behavior for
form factors is not always appropriate.

The pole mass in the expression of $T_1(q^2)$ above obtained by fitting the results calculated by QCD sum rule is
\begin{eqnarray}
m^{T_1}_{{\rm pole}}=5.38\pm 0.23~\mbox{GeV}.
\end{eqnarray}

We have calculated the form factors related to hadronic matrix element $\langle\phi(\varepsilon,p_2)|\bar{s}\gamma_\nu (1-\gamma_5)b|\bar{B}_s^0(p_1)\rangle$ in Ref. \cite{PYQ}, and the results are shown in Table \ref{Ta4}.
\begin{table}[h]
\caption{Form factors related to $\langle\phi(\varepsilon,p_2)|\bar{s}\gamma_\nu (1-\gamma_5)b|\bar{B}_s^0(p_1)\rangle$}
\begin{center}
\begin{tabular}{|c|c|c|c|c|}
\hline
 &$A_0(q^2)$ & $A_1(q^2)$ &$A_2(q^2)$ &$V(q^2)$ \\
                         \hline
$~q^2=0~$  & $0.30\pm0.25 $ & $~0.32\pm0.07~$ & $0.30\pm0.07$ & $0.45\pm0.10$
\\ \hline
$~q^2\neq0~$  & $A_0(q^2)=\frac{A_0(0)}{1-q^2/(m_{{\rm pole}}^{A_0})^2} $&$A_1(0)$ &
$A_2(q^2)=\frac{A_2(0)}{1-q^2/(m_{{\rm pole}}^{A_2})^2}$ &
$V(q^2)=\frac{V(0)}{1-q^2/(m_{{\rm pole}}^V)^2}$ \\ \hline
$m_{{\rm pole}}$  & $5.62\pm 2.38$ GeV & $-$ & $9.20\pm 0.40$ GeV & $5.59\pm 0.27$ GeV \\ \hline
\end{tabular}\end{center}
\label{Ta4}
\end{table}

Next we shall use all of the $\bar{B}_s^0\to \phi$ transition form factors $V$, $A_0$, $A_1$, $A_2$ and $T_1$, $T_2$, $T_3$ calculated by QCD sum rules to investigate the differential decay widths and branching ratios of $\bar{B}_s^0\to \phi l^+ l^-$ decays. The expression of differential decay width is given as \cite{B/Vll},
\begin{eqnarray}
\frac{{\rm d}\Gamma(\bar{B}^0_s\to \phi l^+l^-)}{{\rm d}\hat{s}}
&=&\frac{G^2_F\alpha^2m_{B_s}^5}{2^{10}\pi^5}|V^*_{ts}V_{tb}|^2~\hat{u}(\hat{s})\nonumber\\
&&\times\left\{\frac{|A|^2}{3}\hat{s}\lambda(1+2\frac{ \hat{m}^2_l}{\hat{s}})
+|E|^2\hat{s} \frac{\hat{u}(\hat{s})^2}{3}\right.\nonumber\\
&&+\frac{1}{4\hat{m}^2_{\phi}}\Big[|B|^2\Big(\lambda-\frac{\hat{u}(\hat{s})^2}{3}
+8\hat{m}^2_{\phi}(\hat{s}+2\hat{m}^2_l)\Big) \nonumber\\
&&+|F|^2\Big(\lambda-\frac{\hat{u}(\hat{s})^2}{3}+8\hat{m}^2_{\phi}(\hat{s}-4\hat{m}^2_l)\Big)\Big]\nonumber\\
&&+\frac{\lambda}{4\hat{m}^2_{\phi}}\Big[|C|^2(\lambda-\frac{\hat{u}(\hat{s})^2}{3})+
|G|^2(\lambda-\frac{\hat{u}(\hat{s})^2}{3}+4\hat{m}^2_l(2+2\hat{m}^2_{\phi}-\hat{s})\Big)\Big]\nonumber\\
&&-\frac{1}{2\hat{m}^2_{\phi}}\Big[Re(BC^*)(\lambda-\frac{\hat{u}(\hat{s})^2}{3})(1-\hat{m}^2_{\phi}-\hat{s})\nonumber\\
&&~~~~~~~~~+Re(FG^*)\Big((\lambda-\frac{\hat{u}(\hat{s})^2}{3})(1-\hat{m}^2_{\phi}
-\hat{s})+4\hat{m}^2_l\lambda\Big)\Big]\nonumber\\
&&\left.-2\frac{\hat{m}^2_l}{\hat{m}^2_{\phi}}
\lambda\Big[Re(FH^*)-Re(GH^*)(1-\hat{m}^2_{\phi})\Big]
+\frac{\hat{m}^2_l}{\hat{m}^2_{\phi}}\hat{s}\lambda|H|^2 \right\},
\label{dwidth}
\end{eqnarray}
where $s=q^2$, $\hat{s}=s/m_{B_{s}}^2$, $\hat{m}_q=m_q/m_{B_s}$, $\hat{u}(\hat{s})=\sqrt{\lambda(1-4{ \hat{m}^2_l}/{\hat{s}})}$,
$\lambda\equiv\lambda(1,\hat{m}^2_{\phi},\hat{s})=1+\hat{m}^4_{\phi}+\hat{s}^2-2\hat{s}-2\hat{m}^2_{\phi}(1+\hat{s})$, and
the specific expressions of $A(\hat{s})\thicksim H(\hat{s})$ can be found in Ref. \cite{B/Vll}, which are not listed here for brevity.

Considering the possible long-distance (LD) effects and to avoid the contributions of resonances, some cuts around the resonances of $J/\psi$ and $\psi^{'}$ are taken in the physical distribution of $q^2$. We use the same cuts as that
used by LHCb Collaboration in Ref. \cite{15AQ}.  There are three regions for $\bar{B}_s^0\to \phi e^+ e^-$ and $\bar{B}_s^0\to \phi \mu^+ \mu^-$ decays:
\begin{eqnarray}
\uppercase\expandafter{\mbox{\romannumeral1}}:&&~0.1~{\rm GeV^2}\leq q^2 \leq 8.0~{\rm GeV^2}~;\nonumber  \\
\uppercase\expandafter{\mbox{\romannumeral2}}:&&~11.0~{\rm GeV^2} \leq q^2\leq 12.5~{\rm GeV^2}~;\\
\uppercase\expandafter{\mbox{\romannumeral3}}:&&~15.0~{\rm GeV^2} \leq q^2 \leq 19.0~{\rm GeV^2}~.\nonumber
\end{eqnarray}
and two regions for $\bar{B}_s^0\to \phi \tau^+ \tau^-$ decay:
\begin{eqnarray}
\uppercase\expandafter{\mbox{\romannumeral1}}:&&~11.0~{\rm GeV^2} \leq q^2\leq 12.5~{\rm GeV^2}~;\nonumber \\
\uppercase\expandafter{\mbox{\romannumeral2}}:&&~15.0~{\rm GeV^2} \leq q^2 \leq 19.0~{\rm GeV^2}~.
\end{eqnarray}

\begin{figure}[tbp]
  \centering
  \includegraphics[width=0.95\textwidth,origin=l,angle=0]{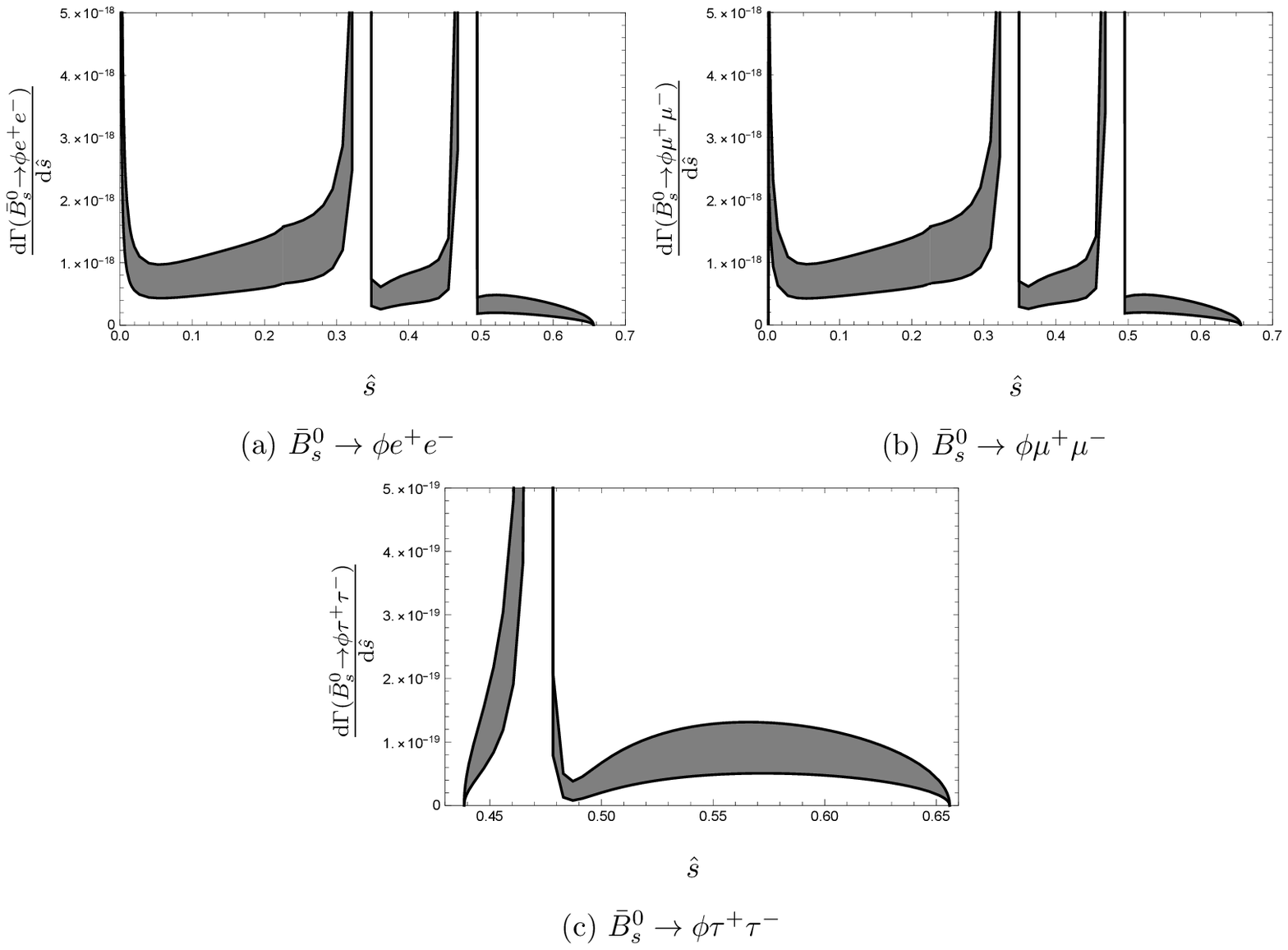}
  \caption{The differential decay widths of $\bar{B}_s^0\to \phi l^+ l^-$ ($l=e, \mu, \tau$) on $q^2$ with LD effects. The grey bands denote the relevant uncertainties.} \label{q2}
\end{figure}
The $q^2$-dependence of differential decay widths with long-distance (LD) effects are shown in Fig.\ref{q2},
where the grey bands denote the relevant uncertainties.
Integrating the differential decay width in Eq.~(\ref{dwidth}) with respect to $q^2$ within the relevant region,
we can obtain the value of integrated decay width $\Gamma(\bar{B}^0_s\to \phi l^+l^-)$.
According to the definition of decay branching ratio
\begin{eqnarray}
\mbox{Br}(\bar{B}^0_s\to \phi l^+l^-)=\frac{\Gamma(\bar{B}^0_s\to \phi l^+l^-)}{\Gamma_{{\rm total}}(\bar{B}^0_s)},
\end{eqnarray}
and the total decay width of $\bar{B}_s^0$ meson: $\Gamma_{{\rm total}}(\bar{B}^0_s)=4.362\times10^{-13} \mbox{GeV} $ \cite{PDG2018},
we can get the branching ratios of the three semileptonic decay channels of $\bar{B}_s^0\to \phi l^+ l^-$ ($l=e, \mu, \tau$),
\begin{eqnarray}
\mbox{Br}(\bar{B}_s^0\to \phi e^+e^-)=(7.12\pm 1.40)\times10^{-7},
\end{eqnarray}
\begin{eqnarray}\label{mu}
\mbox{Br}(\bar{B}_s^0\to \phi \mu^+\mu^-)=(7.06\pm 1.59)\times10^{-7},
\end{eqnarray}
\begin{eqnarray}
\mbox{Br}(\bar{B}_s^0\to \phi \tau^+\tau^-)=(3.49\pm 1.69)\times10^{-8}.
\end{eqnarray}
The experimental result of the total branching ratio of $\bar{B}_s^0\to \phi \mu^+ \mu^-$ is \cite{15AQ}
\begin{equation}\label{expmu}
\mbox{Br}(\bar{B}_s^0\to \phi \mu^+ \mu^-)=(7.97^{+0.45}_{-0.43}\pm0.22\pm0.23\pm0.60)\times10^{-7}.
\end{equation}
We find agreement between our predictions and the experimental data within uncertainties.

Furthermore, in order to show the physical effects caused by the sign of the form factors,
we change the sign of the form factors $V$, $A_1$, $T_1$ and $T_3$ as that of Ref. \cite{R.K-SumR} to calculate the
branching ratio of $\bar{B}_s^0\to \phi \mu^+ \mu^-$ again, and obtain the central value of the branching ratio of
as follows
\begin{eqnarray}
\mbox{Br}(\bar{B}_s^0\to \phi \mu^+\mu^-)=6.14\times10^{-6}. \label{minus/mu}
\end{eqnarray}
From Eq.~(\ref{minus/mu}) we can find that the branching ratio of $\bar{B}_s^0\to \phi \mu^+\mu^-$ calculated in this way is nearly an order of magnitude larger than the experimental data in Eq.~(\ref{expmu}). So the physical effect of
the sign of the form factors are crucial.


\section{Summary}
We revisit the semi-leptonic decays of $\bar{B}_s^0\to \phi l^+ l^-$ ($l=e, \mu, \tau$)
with QCD sum rule method. The $\bar{B}_s^0\to \phi$ transition form factors
$V$, $A_0$, $A_1$, $A_2$ \cite{PYQ} and $T_1$, $T_2$, $T_3$ are calculated,
then they are used to obtain the branching ratios of $\bar{B}_s^0\to \phi e^+ e^-$,
$\bar{B}_s^0\to \phi\mu^+ \mu^-$ and $\bar{B}_s^0\to \phi \tau^+ \tau^-$ respectively.
For the measured decay channel $\bar{B}_s^0\to \phi\mu^+ \mu^-$, our theoretical result is
$\mbox{Br}(\bar{B}_s^0\to \phi \mu^+\mu^-)=(7.06\pm 1.59)\times10^{-7}$,
which is well consistent with the latest experimental data $\mbox{Br}(\bar{B}_s^0\to \phi \mu^+\mu^-)=(7.97^{+0.45}_{-0.43}\pm0.22\pm0.23\pm0.60)\times10^{-7}$ from
LHCb Collaboration within uncertainties. For the unmeasured decay channels:
$\bar{B}_s^0\to \phi e^+ e^-$ and $\bar{B}_s^0\to \phi\tau^+ \tau^-$, we
hope that our theoretical predictions are useful for experimental test in the future.

\acknowledgments

This work is supported in part by the National Natural Science
Foundation of China under Contracts No. 11875168 and No. 11375088.

\newpage
\begin{center}{\bf Appendix A}\end{center}
The explicit form of the relevant Borel transformed Coefficients $ \hat{B}\kappa_0$, $ \frac{1}{2}\hat{B}(\kappa_1-\kappa_3)$ and $ \hat{B}\kappa_5$
in Eq.~(\ref{14}) are given in the following.

\noindent 1) Results for Borel transformed $\kappa_0$:

$$ \hat{B}\kappa_0= \hat{B}\kappa_0^{{\rm pert}}+ \hat{B}\kappa_0^{(3)}+\hat{B}\kappa_0^{(4)}+
\hat{B}\kappa_0^{(5)}+ \hat{B}\kappa_0^{(6)}\; ,$$
where
$$
\begin{array}{ll}
 \hat{B}\kappa_0^{{\rm pert}}~=&\int^{s_2^0}_{4m_s^2}
 {\rm d}s_2\int^{s_1^0}_{s_1^L}{\rm d}s_1
\displaystyle
\frac{3{\rm e}^{-s_1/M_1^2-s_2/M_2^2}}{8\pi^2\lambda^{3/2} M_1^2 M_2^2}\left[-2 \lambda  m_b m_s+4 s_2 m_b^2 m_s^2\right. \\
&\left.-2 s_2 m_b^4+s_2(-\lambda -2 m_s^4+q^4-2 q^2s_2+s_1^2-2s_1 s_2+s_2^2)\right],
\end{array}
\eqno(A1)$$
with $\lambda=(s_1+s_2-q^2)^2-4s_1 s_2$.

$$
\begin{array}{ll}
 \hat{B}\kappa_0^{(3)}~=&\displaystyle
\frac{{\rm e}^{-m_b^2/M_1^2-m_s^2/M_2^2}}{6M_1^8 M_2^8}
[M_2^2 m_b^3 m_s^2 (M_1^2+M_2^2)(3M_1^2 M_2^2-m_s^2 (M_1^2+M_2^2))\\
&+M_1^2 M_2^2 m_b^2 m_s(M_2^2 m_s^2(3 M_1^2+3 M_2^2+q^2)-2 m_s^4(M_1^2+M_2^2)-3M_1^2 M_2^4)\\
&-M_2^4 m_b^4 m_s^3(M_1^2+M_2^2)+M_1^2 m_b (M_2^2 m_s^4(M_1^2+M_2^2)(4 M_1^2+M_2^2+q^2)\\
&-m_s^6(M_1^2+M_2^2)^2-3M_1^2 M_2^4 m_s^2(M_1^2+M_2^2+q^2)+6 M_1^4 M_2^6)+M_1^4 m_s\\
&\times(M_2^4 m_s^2(3 M_1^2+2 q^2)+M_2^2 m_s^4(3 M_1^2+3 M_2^2+q^2)-m_s^6 (M_1^2+M_2^2)\\
&+3 M_1^2M_2^6)]
\times\langle \bar{s}s\rangle\; ,
 \end{array}
\eqno(A2)
$$

$$
\begin{array}{ll}
 \hat{B}\kappa_0^{(4)}~=&-\int^{s_2^0}_{4m_s^2}
 {\rm d}s_2\int^{s_1^0}_{s_1^L}{\rm d}s_1
\displaystyle
\frac{{\rm e}^{-s_1/M_1^2-s_2/M_2^2}}{96 \pi^2 \lambda^{3/2}M_1^2 M_2^2}(-4q^2+5s_1+4s_2)
\times 4 \pi\alpha_s\langle GG\rangle\; ,
\end{array}
\eqno(A3)$$

$$
\begin{array}{ll}
 \hat{B}\kappa_0^{(5)}~=&-\displaystyle
\frac{{\rm e}^{-m_b^2/M_1^2-m_s^2/M_2^2}}{12M_1^8 M_2^8}
[M_1^2 M_2^2 m_b^2 m_s (M_2^2 (3M_1^2+q^2)-2 m_s^2(M_1^2+M_2^2))\\
&+M_2^2 m_b^3 (M_1^2+M_2^2) (3M_1^2 M_2^2-m_s^2 (M_1^2+M_2^2))+M_1^2 m_b (M_2^2 m_s^2(M_1^2\\
&+M_2^2)(5 M_1^2-2 M_2^2+q^2)-m_s^4 (M_1^2+M_2^2)^2+M_1^2 M_2^4 (M_1^2-6 M_2^2-3q^2))\\
&-M_2^4 m_b^4 m_s(M_1^2+M_2^2)+M_1^4 m_s(M_2^2 m_s^2(4M_1^2+M_2^2+q^2)-m_s^4(M_1^2+M_2^2)\\
&-M_2^4(-2 M_1^2+3M_2^2+q^2))]
\times g\langle \bar{s}\sigma TG s \rangle\; ,
\end{array}
\eqno(A4)
$$

$$
\begin{array}{ll}
 \hat{B}\kappa_0^{(6)}~=&\displaystyle
 -\frac{{\rm e}^{-m_b^2/M_1^2-m_s^2/M_2^2}} {81 M_1^8 M_2^8 m_s^3(m_b^2-q^2)}
[M_2^2 m_b^5 m_s^4 (M_1^2+M_2^2)^2+M_2^4 m_b^6 m_s^3 (M_1^2+M_2^2)\\
&+M_2^2 m_b^4 m_s^3 (2M_1^2 m_s^2 (M_1^2+M_2^2)-M_2^2(-3 M_1^4+M_1^2(15M_2^2+2 q^2)+M_2^2 q^2))\\
&+M_1^2 m_b^2 m_s(18M_1^4M_2^6 ({\rm e}^{\frac{m_s^2}{M_2^2}}-1)+M_1^2 m_s^6 (M_1^2+M_2^2)-M_2^2 m_s^4(M_1^4+3M_1^2\\
&\times(3M_2^2+q^2)+2M_2^2 q^2)+M_2^4 m_s^2 (50M_1^4-4M_1^2 (6M_2^2+q^2)+q^2 (15M_2^2+q^2)))\\
&+M_1^2 m_b (54M_1^4 M_2^6 q^2(e^{\frac{m_s^2}{M_2^2}}-1)-18M_1^4 M_2^4 m_s^2 (M_2^2 ({\rm e}^{\frac{m_s^2}{M_2^2}}-1)+3 q^2)-q^2 m_s^6 \\
&\times(M_1^2+M_2^2)^2+M_2^2 q^2 m_s^4 (M_1^4+M_1^2 (20M_2^2+q^2)+M_2^2 (13M_2^2+q^2)))-m_b^3\\
&\times(-54 M_1^6 M_2^4 m_s^2+54 M_1^6 M_2^6 ({\rm e}^{\frac{m_s^2}{M_2^2}}-1)-M_1^2 m_s^6 (M_1^2+M_2^2)^2+M_2^2 m_s^4 (M_1^6\\
&+2 M_1^4(10M_2^2+q^2)+M_1^2(13 M_2^4+3 M_2^2 q^2)+M_2^4 q^2))+M_1^4 q^2 m_s (M_2^2 m_s^4(M_1^2\\
&+9 M_2^2+q^2)-m_s^6(M_1^2+M_2^2)+18 M_1^2 M_2^6 ({\rm e}^{\frac{m_s^2}{M_2^2}}-1)+M_2^4 m_s^2(-50 M_1^2+24 M_2^2\\
&+q^2))]
\times g^2\langle \bar{s}s\rangle^2\; .
\end{array}
\eqno(A5)
$$

In the perturbative diagram, we consider the condition that all internal quarks are on their mass shell \cite{R1},
which gives the lower limit of the integration $s_1^L$ as
$$ s_1^L=\frac{m_b^2}{m_b^2-q^2}s_2+m_b^2,$$

\noindent 2) Results for Borel transformed $(\kappa_1-\kappa_3)$ :

$$ \frac{1}{2}\hat{B}(\kappa_1-\kappa_3)= \frac{1}{2}\hat{B}\kappa_-^{{\rm pert}}+ \frac{1}{2}\hat{B}\kappa_-^{(3)}+\frac{1}{2}\hat{B}\kappa_-^{(4)}+
\frac{1}{2}\hat{B}\kappa_-^{(5)}+ \frac{1}{2}\hat{B}\kappa_-^{(6)}\; ,$$

$$
\begin{array}{ll}
 \frac{1}{2}\hat{B}\kappa_-^{{\rm pert}}~=&-\int^{s_2^0}_{4m_s^2}
 {\rm d}s_2\int^{s_1^0}_{s_1^L}{\rm d}s_1
\displaystyle
\frac{3{\rm e}^{-s_1/M_1^2-s_2/M_2^2}}{8 \pi^2 \lambda^{5/2}M_1^2 M_2^2}
[s_2 m_b^4 (2 \lambda -3 q^4+12 q^2 s_2+3s_1^2+6s_1 s_2\\
&-9s_2^2)-2s_2 m_b^2 (m_s^2 (2 \lambda -3 q^4+12 q^2 s_2+3s_1^2+6s_1 s_2-9 s_2^2)-q^6\\
&-2 q^4 (s_1-3s_2)+q^2 (\lambda +s_1^2+8s_1s_2-9s_2^2)+2 (s_1^3-3s_1 s_2^2+2s_2^3\\
&-\lambda s_2))+\lambda^2 m_b m_s+s_2 m_s^4 (2 \lambda -3 q^4+12 q^2s_2+3s_1^2+6s_1 s_2-9s_2^2)\\
&+m_s^2(-\lambda ^2-2 q^6s_2-q^4(\lambda +4s_1 s_2-12s_2^2)+2 q^2s_2(3 \lambda +s_1^2+8s_1s_2\\
&-9 s_2^2)+4 s_1^3 s_2+\lambda s_1^2+2s_1 s_2 (\lambda -6 s_2^2)+8 s_2^4-7 \lambda  s_2^2)+s_2(Q^3 (s_2\\
&-2 s_1)-q^4(s_1^2-8s_1 s_2+3s_2^2)+q^2 (s_1-s_2)(\lambda +2 s_1^2+5 s_1 s_2-3 s_2^2)\\
&+(s_1^2-s_2^2)(-\lambda +s_1^2-2 s_1s_2+s_2^2))] ,
\end{array}
\eqno(A6)
$$

$$
\begin{array}{ll}
 \frac{1}{2}\hat{B}\kappa_-^{(3)}~=&-\displaystyle
\frac{{\rm e}^{-m_b^2/M_1^2-m_s^2/M_2^2}} {12M_1^8 M_2^8}
[-M_2^2 m_b^3 m_s^2 (M_1^2+M_2^2)(m_s^2(M_2^2-M_1^2)+3 M_1^2\\
&\times M_2^2)+M_1^2 M_2^4 m_b^2 m_s (m_s^2 (5 M_1^2+M_2^2-q^2)+3 M_1^2 M_2^2)+M_2^4\\
&\times m_b^4 m_s^3(M_1^2+M_2^2)+M_1^2 m_b (m_s^6 (M_1^4-M_2^4)+3 M_1^2 M_2^4 m_s^2 (M_1^2\\
&+M_2^2+q^2)+M_2^2 m_s^4(-4 M_1^4-M_1^2 (3 M_2^2+q^2)+M_2^2(M_2^2+q^2))\\
&-6 M_1^4 M_2^6)+M_1^4 m_s(-M_2^4 m_s^2(9 M_1^2+4 q^2)+M_2^2 m_s^4(7 M_1^2+3M_2^2\\
&+q^2)-m_s^6(M_1^2+M_2^2)+9 M_1^2 M_2^6)]
\times\langle \bar{s}s\rangle\; ,
 \end{array}
\eqno(A7)
$$

$$
\begin{array}{ll}
 \frac{1}{2}\hat{B}\kappa_-^{(4)}~=&\int^{s_2^0}_{4m_s^2}
 {\rm d}s_2\int^{s_1^0}_{s_1^L}{\rm d}s_1
\displaystyle
\frac{{\rm e}^{-s_1/M_1^2-s_2/M_2^2}}{96 \pi^2 \lambda^{5/2} M1^2 M2^2}
[3 q^6-3q^4 (5s_1+3s_2)+q^2(-5 \lambda \\
&+21s_1^2+22 s_1 s_2+9 s_2^2)-9s_1^3-13s_1^2 s_2+4 \lambda s_1+25s_1 s_2^2-3s_2^3\\
&+9 \lambda s_2]
\times 4 \pi\alpha_s\langle GG\rangle\; ,
\end{array}
\eqno(A8)$$

$$\begin{array}{ll}
  \frac{1}{2}\hat{B}\kappa_+^{(5)}~=&-\displaystyle
\frac{{\rm e}^{-m_b^2/M_1^2-m_s^2/M_2^2}}{24 M_1^8 M_2^8}
[M_2^2 m_b^3 (M_1^2+M_2^2) (m_s^2(M_2^2-M_1^2)+3M_1^2 M_2^2)\\
&+M_1^2 M_2^4 m_b^2 m_s(-7M_1^2-4 M_2^2+q^2)-M_2^4 m_b^4 m_s (M_1^2+M_2^2)+M_1^2 \\
&\times m_b(m_s^4(M_2^4-M_1^4)+M_2^2 m_s^2(5M_1^4+M_1^2(q^2-M_2^2)+2M_2^4-M_2^2 \\
&\times q^2)+M_1^2 M_2^4 (M_1^2-3 (4M_2^2+q^2)))+M_1^4 m_s(-M_2^2 m_s^2(8 M_1^2+M_2^2\\
&+q^2)+m_s^4(M_1^2+M_2^2)+M_2^4 (8M_1^2+3M_2^2+5 q^2))]
\times g\langle
\bar{s}\sigma TG s \rangle\; ,
\end{array}
\eqno(A9)$$

$$
\begin{array}{ll}
  \frac{1}{2}\hat{B}\kappa_+^{(6)}~=&-\displaystyle
\frac{{\rm e}^{-m_b^2/M_1^2-m_s^2/M_2^2}}{162M_1^8 M_2^8 m_s^3(m_b^2-q^2)}
[M_2^2 m_b^5 m_s^4(M_1^4-M_2^4)+M_2^4 m_b^6 m_s^3(M_1^2\\
&+M_2^2)-M_2^4 m_b^4 m_s^3(-5M_1^4+M_1^2(17M_2^2+2 q^2)+M_2^2 q^2)+M_1^2\\
&\times m_b^2 m_s(54 M_1^4 M_2^6({\rm e}^{\frac{m_s^2}{M_2^2}}-1)+M_1^2 M_2^2 m_s^4(5M_1^2+9M_2^2+q^2)\\
&-M_1^2 m_s^6(M_1^2+M_2^2)+M_2^4 m_s^2 (-44 M_1^4-6 M_1^2(4 M_2^2+q^2)+q^2\\
&\times(17M_2^2+q^2)))+M_1^2 m_b (q^2 m_s^6(M_2^4-M_1^4)+54 M_1^4 M_2^6 q^2 ({\rm e}^{\frac{m_s^2}{M_2^2}}-1)\\
&-18M_1^4 M_2^4 m_s^2 (M_2^2({\rm e}^{\frac{m_s^2}{M_2^2}}-1)+3 q^2)+M_2^2 q^2 m_s^4(M_1^4+M_1^2 (18 M_2^2\\
&+q^2)-M_2^2(13 M_2^2+q^2)))+m_b^3(54M_1^6 M_2^4 m_s^2-54 M_1^6 M_2^6({\rm e}^{\frac{m_s^2}{M_2^2}}-1)\\
&+m_s^6 (M_1^6-M_1^2 M_2^4)+M_2^2 m_s^4(-M_1^6-2 M_1^4 (9 M_2^2+q^2)+M_1^2M_2^2\\
&\times(13M_2^2+q^2)+M_2^4 q^2))+M_1^4 m_s(-M_2^2 q^2 m_s^4(5 M_1^2+9 M_2^2+q^2) \\
&+q^2 m_s^6(M_1^2+M_2^2)-54 M_1^2 M_2^6 q^2 ({\rm e}^{\frac{m_s^2}{M_2^2}}-1)+M_2^4 m_s^2 (M_1^2(44q^2-72 \\
&\times M_2^2)+q^2 (24 M_2^2+q^2)))]
 \times g^2\langle \bar{s}s\rangle^2\; .
\end{array}
\eqno(A10)$$

\noindent 3) Results for Borel transformed $\kappa_5$ :

$$
\begin{array}{ll}
 \hat{B}\kappa_5^{{\rm pert}}~=&-\int^{s_2^0}_{4m_s^2}
 {\rm d}s_2\int^{s_1^0}_{s_1^L}{\rm d}s_1
\displaystyle
\frac{3{\rm e}^{-s_1/M_1^2-s_2/M_2^2}}{8 \pi^2 \lambda^{3/2} M_1^2 M_2^2}
[-s_2 m_b^2 (-\lambda +2 m_s^2 (q^2+s_1-s_2)\\
&+q^4+2 q^2 (s_1-s_2)+s_1^2-2s_1 s_2+s_2^2)-\lambda  m_b m_s (q^2-s_1+s_2)\\
&+s_2 m_b^4 (q^2+s_1-s_2)+m_s^2 (q^4 s_2+2 q^2 (\lambda +s_1 s_2-s_2^2)+s_2(-\lambda +s_1^2\\
&-2 s_1 s_2+s_2^2))+s_2 m_s^4 (q^2+s_1-s_2)+q^2 s_1 s_2 (q^2+s_1-s_2)]
 \; ,
\end{array}
\eqno(A11)
$$

$$
\begin{array}{ll}
 \hat{B}\kappa_5^{(3)}~=&\displaystyle
\frac{{\rm e}^{-m_b^2/M_1^2-m_s^2/M_2^2}}{12 M_1^8 M_2^8}
[M_2^2 m_b^5 m_s^2 (M_1^2+M_2^2) (3 M_1^2 M_2^2-m_s^2(M_1^2\\
&+M_2^2))-M_2^4 m_b^6 m_s^3(M_1^2+M_2^2)+M_1^2 m_b^2 m_s(M_2^2 m_s^4 (M_1^2-M_2^2)\\
&\times(5 M_1^2+5 M_2^2+q^2)-M_2^4 m_s^2(6 M_1^4+M_1^2 q^2-2 M_2^2 q^2+q^4)+m_s^6\\
&\times(-M_1^4+M_1^2 M_2^2+2 M_2^4)+3 M_1^2 M_2^6(2 M_1^2+q^2))+M_1^2 m_b (-m_s^6\\
&\times(M_1^2+M_2^2) (M_1^2(5 M_2^2-q^2)+2 M_2^2 (M_2^2+q^2))+m_s^8(M_1^2+M_2^2)^2\\
&+3 M_1^2M_2^4 m_s^2(q^4-2 M_1^2(M_2^2-q^2))+M_2^2 m_s^4(M_1^4(6 M_2^2-5 q^2)\\
&+M_1^2(6 M_2^4-q^4)+M_2^2 q^2(2M_2^2+q^2))-6M_1^4 M_2^6 q^2)+m_b^4(-3 M_1^4\\
&\times M_2^6 m_s+M_2^4 m_s^3 (5 M_1^4+M_1^2(5 M_2^2+2 q^2)+M_2^2 q^2)+m_s^5(-2 M_1^4 M_2^2\\
&-M_1^2 M_2^4+M_2^6))+m_b^3(-m_s^6 (M_1^2-M_2^2)(M_1^2+M_2^2)^2-M_2^2 m_s^4(M_1^2\\
&+M_2^2)(-5M_1^4+M_1^2 (M_2^2-2 q^2)+M_2^2 q^2)-3 M_1^2 M_2^4 m_s^2 (2 M_1^4+2 M_1^2\\
&\times(M2^2+q^2)+M_2^2 q^2)+6 M_1^6 M_2^6)+M_1^4 m_s (m_s^8(M_1^2+M_2^2)+M_2^2 m_s^4\\
&\times(6M_1^2 (M_2^2+q^2)+3 M_2^4+9 M_2^2 q^2+q^4)-m_s^6(M_1^2(5 M_2^2+q^2)+5 M_2^4\\
&+2 M2^2 q^2)-M_2^4 m_s^2 (M_1^2(6M_2^2+5 q^2)+4 q^2 (M_2^2+q^2))+9M_1^2 M_2^6 q^2)]\\
& \times\langle \bar{s}s\rangle\; ,
 \end{array}
\eqno(A12)
$$

$$
\begin{array}{ll}
 \hat{B}\kappa_5^{(4)}~=&-\int^{s_2^0}_{4m_s^2}
 {\rm d}s_2\int^{s_1^0}_{s_1^L}{\rm d}s_1
\displaystyle
\frac{{\rm e}^{-s_1/M_1^2-s_2/M_2^2}}{192 \pi^2 \lambda^{3/2} M_1^2 M_2^2}
[-5 \lambda +q^4-2 q^2 (3 s_1+s_2)+5s_1^2\\
&-6 s_1 s_2+s_2^2]
\times 4 \pi\alpha_s\langle GG\rangle\; ,
\end{array}
\eqno(A13)$$

\newpage
$$
\begin{array}{ll}
 \hat{B}\kappa_5^{(6)}~=&\displaystyle
\frac{{\rm e}^{-m_b^2/M_1^2-m_s^2/M_2^2}}{162 M_1^8 M_2^8(m_b^2-q^2) m_s^3}
[-M_2^4(M_1^2+M_2^2) m_s^3 m_b^8-M_2^2(M_1^2+M_2^2)^2\\
&\times m_s^4 m_b^7+M_2^2 m_s^3((-M_1^4+(17 M_2^2+3 q^2) M_1^2+2 M_2^2 q^2) M_2^2+(-2 M_1^4\\
&-M_2^2 M_1^2+M_2^4) m_s^2) m_b^6+(54(-1+{\rm e}^{\frac{m_s^2}{M_2^2}}) M_2^6 M_1^6-54 M_2^4 m_s^2 M_1^6+M_2^2 \\
&\times(2 M_1^4+(22 M_2^2+3 q^2) M_1^2+14 M_2^4+3 M_2^2 q^2) m_s^4 M_1^2-(M_1^2-M_2^2)\\
&\times(M_1^2+M_2^2)^2 m_s^6) m_b^5-m_s(18(-1+{\rm e}^{\frac{m_s^2}{M_2^2}}) M_1^6 M_2^6+(53 M_1^6+(9 M_2^2\\
&-3 q^2) M_1^4+3 q^2 (11 M_2^2+q^2) M_1^2+M_2^2 q^4) m_s^2 M_2^4+(-3 M_1^6-3 (4 M_2^2\\
&+q^2) M_1^4+17 M_2^4 M_1^2+M_2^4 q^2) m_s^4 M_2^2+(M_1^6-M_2^2 M_1^4-2 M_2^4 M_1^2) m_s^6) \\
&\times m_b^4+(M_1^2 (M_1^2+M_2^2)^2 m_s^8-(2 (M_2^2-q^2) M_1^6+22 M_2^4 M_1^4+(14 M_2^6\\
&+3 q^2 M_2^4) M_1^2+M_2^6 q^2) m_s^6+M_2^2 ((54 M_2^2-4 q^2) M_1^6-(60 M_2^4+40 q^2 M_2^2\\
&+3 q^4) M_1^4+M_2^4 q^4) m_s^4+18 M_1^6 M2^4 ((-1+{\rm e}^{\frac{m_s^2}{M_2^2}}) M_2^2+6 q^2) m_s^2-108 \\
&\times(-1+e^{\frac{m_s^2}{M_2^2}}) M_1^6 M_2^6 q^2) m_b^3+M_1^2 m_s (M_1^2 (M_1^2+M_2^2) m_s^8-M_2^2 (3 M_1^4+\\
&(11M_2^2+3 q^2) M_1^2+2 M_2^2 q^2) m_s^6+(9 M_1^2 M_2^6+q^2 (17 M_2^2+q^2) M_2^4+M_1^4 \\
&\times(49 M_2^2+q^2) M_2^2) m_s^4+M_2^4 ((4 q^2-98 M_2^2) M_1^4-q^2 (16 M_2^2+3 q^2) M_1^2\\
&+q^4 (16M_2^2+q^2)) m_s^2+72(-1+{\rm e}^{\frac{m_s^2}{M_2^2}}) M_1^4 M_2^6 q^2) m_b^2+M_1^2 (-(M_1^2+M_2^2)^2 \\
&q^2 m_s^8+q^2((2 M_2^2-q^2) M_1^4+M_2^2 (22 M_2^2+q^2) M_1^2+2 M_2^4 (7 M_2^2+q^2)) m_s^6\\
&+M_2^2(2 (9 M_2^4-27 q^2 M_2^2+q^4) M_1^4+q^2 (60 M_2^4+18 q^2 M_2^2+q^4) M_1^2-M_2^2 q^4 \\
&\times(14 M_2^2+q^2)) m_s^4-18 M_1^4 M_2^4 q^2((-1+{\rm e}^{\frac{m_s^2}{M_2^2}}) M_2^2+3 q^2) m_s^2+54 (-1+{\rm e}^{\frac{m_s^2}{M_2^2}})\\
&\times M_1^4 M_2^6 q^4) m_b+M_1^4 q^2 m_s (-(M_1^2+M_2^2) m_s^8+(11 M_2^4+2 q^2 M_2^2+M_1^2 \\
&\times(3 M_2^2+q^2)) m_s^6-M_2^2 (9 M_2^4+12 q^2 M_2^2+q^4+M_1^2 (49 M_2^2+4 q^2)) m_s^4\\
&+M_2^4 ((26 M_2^2+49 q^2) M_1^2+q^2 (25 M_2^2+q^2)) m_s^2-54(-1+{\rm e}^{\frac{m_s^2}{M_2^2}}) M_1^2 M_2^6 q^2)]\\
&
\times g^2\langle \bar{s}s\rangle^2\; .
\end{array}
\eqno(A14)
$$

\newpage
$$
\begin{array}{ll}
  \hat{B}\kappa_5^{(5)}~=&\displaystyle
\frac{{\rm e}^{-m_b^2/M_1^2-m_s^2/M_2^2}}{24 M_1^8 M_2^8}
[-M_2^2 m_b^5(M_1^2+M_2^2)(3 M_1^2 M_2^2-m_s^2(M_1^2+M_2^2))\\
&+M_2^4 m_b^6 m_s (M_1^2+M_2^2)+M_1^2 m_b^2 m_s(M_2^2 m_s^2(-6 M_1^4+2 M_1^2 M_2^2\\
&-M_1^2 q^2+2 M_2^4+M_2^2 q^2)+m_s^4 (M_1^4-M_1^2 M_2^2-2 M_2^4)+M_2^4 (M_1^4\\
&-5 M_2^2 q^2+q^4))-M_1^2 m_b(m_s^6(M_1^2+M_2^2)^2-m_s^4 (M_1^2+M_2^2) (M_1^2\\
&\times(6 M_2^2-q^2)-M_2^4+2 M_2^2 q^2)+M_2^2 m_s^2(6M_1^4 (M_2^2-q^2)+M_1^2(13\\
&\times M_2^4+4 M_2^2 q^2-q^4)+M_2^2 q^2(q^2-M_2^2))+M_1^2 M_2^4 (M_1^2(7 M_2^2+2 q^2)\\
&+3 q^2(3M_2^2+q^2)))-M_2^2 m_b^4 m_s(m_s^2(-2M_1^4-M_1^2M_2^2+M_2^4)+M_2^2\\
&\times(5M_1^4+2 M_1^2 (M_2^2+q^2)+M_2^2 q^2))+m_b^3 (m_s^4 (M_1^2-M_2^2) (M_1^2+M_2^2)^2\\
&+M2^2 m_s^2(M_1^2+M_2^2) (-6M_1^4+M_1^2 (4M_2^2-2 q^2)+M_2^2 q^2)+M_1^2 M_2^4\\
 &\times(2M_1^4+M_1^2 (9M_2^2+6 q^2)+3M_2^2 q^2))-M_1^4 m_s (m_s^6 (M_1^2+M_2^2)+M_2^2\\
&\times  m_s^2 (M_1^2 (5 M_2^2+7 q^2)+4M_2^4+8M_2^2 q^2+q^4)-m_s^4 (M_1^2(6 M_2^2+q^2)\\
&+3M_2^4+2M_2^2 q^2)+M_2^4(M_1^2(5 M_2^2-3 q^2)-q^2(8M_2^2+5 q^2)))]
\times\langle \bar{s}s\rangle\; ,
 \end{array}
\eqno(A15)
$$

\newpage
\begin{center}{\bf Appendix B}\end{center}

As shown in Eqs. (\ref{12}) and (\ref{13}), the Wilson coefficients contributed by the diagrams of Fig.1(a)-(f) are
$\kappa_i^{(4)}$, $i=0,\cdots, 5$. After Borel transformation, they will finally contribute to the form factors.
To show how large numerically the contribution of each diagram in Fig.1 is, we take the Borel transformed Wilson
coefficient $\hat{B}\kappa_0^{(4)}$ as an example. The contributions of Fig.1(a)-(f) are given as
$$
[\hat{B}\kappa_0^{(4)}]_{({\rm a})}=-\int^{s_2^0}_{4m_s^2}
 {\rm d}s_2\int^{s_1^0}_{s_1^L}{\rm d}s_1
\displaystyle
\frac{{\rm e}^{-s_1/M_1^2-s_2/M_2^2}}{96 \pi^2 \lambda^{3/2}M_1^2 M_2^2}(-q^2+3s_1+s_2)
\times 4 \pi\alpha_s\langle GG\rangle\; ,
\eqno{(B1)} $$
$$
 [\hat{B}\kappa_0^{(4)}]_{({\rm b})}=-\int^{s_2^0}_{4m_s^2}
 {\rm d}s_2\int^{s_1^0}_{s_1^L}{\rm d}s_1
\displaystyle
\frac{{\rm e}^{-s_1/M_1^2-s_2/M_2^2}}{96 \pi^2 \lambda^{3/2}M_1^2 M_2^2}(-q^2+s_1+s_2)
\times 4 \pi\alpha_s\langle GG\rangle\; ,
\eqno{(B2)} $$
$$
 [\hat{B}\kappa_0^{(4)}]_{({\rm c})}=-\int^{s_2^0}_{4m_s^2}
 {\rm d}s_2\int^{s_1^0}_{s_1^L}{\rm d}s_1
\displaystyle
\frac{{\rm e}^{-s_1/M_1^2-s_2/M_2^2}}{96 \pi^2 \lambda^{3/2}M_1^2 M_2^2}(-q^2+s_2)
\times 4 \pi\alpha_s\langle GG\rangle\; ,~~~~~~
\eqno{(B3)} $$
$$
 [\hat{B}\kappa_0^{(4)}]_{({\rm e})}=-\int^{s_2^0}_{4m_s^2}
 {\rm d}s_2\int^{s_1^0}_{s_1^L}{\rm d}s_1
\displaystyle
\frac{{\rm e}^{-s_1/M_1^2-s_2/M_2^2}}{96 \pi^2 \lambda^{3/2}M_1^2 M_2^2}(-q^2+s_1+s_2)
\times 4 \pi\alpha_s\langle GG\rangle\; ,
\eqno{(B4)}$$
with $[\hat{B}\kappa_0^{(4)}]_{({\rm d})}=0$, $[\hat{B}\kappa_0^{(4)}]_{({\rm f})}=0$.
The numerical results for the contributions of Fig.1(a)-(f) are
given below by taking a group of typical values of the input parameters as an example. When taking
$$
\begin{array}{ll}
&s_1^0~=35.9\mbox{GeV}^2, ~~~~~s_2^0~=2.1\mbox{GeV}^2,\\
&M_1^2=16.0\mbox{GeV}^2, ~~~~M_2^2=1.8\mbox{GeV}^2, ~~~~q^2=5\mbox{GeV}^2
\end{array}
\eqno{(B5)} $$
for example, the numerical results for $\hat{B}\kappa_0^{(4)}$ are
$$\begin{array}{ll}
&[\hat{B}\kappa_0^{(4)}]_{({\rm a})}=-8.68\times10^{-7}, ~~[\hat{B}\kappa_0^{(4)}]_{({\rm b})}=-2.52\times10^{-7},\\
&[\hat{B}\kappa_0^{(4)}]_{({\rm c})}=~~5.64\times10^{-8}, ~~[\hat{B}\kappa_0^{(4)}]_{({\rm e})}=-2.52\times10^{-7},
\end{array}
\eqno{(B6)} $$
which are very small compared to Wilson coefficients contributed by other diagrams. For example, the numerical
result for $\hat{B}\kappa_0^{(3)}$, the contribution of quark-quark condensate, is
$$\hat{B}\kappa_0^{(3)}= -5.35\times 10^{(-4)}    \eqno{(B7)}$$
by taking the same values for input parameters. The smallness of the gluon condensate contributions implies 
that they can be neglected in the numerical analysis for the transition form factors. Actually they 
can be viewed as higher order corrections in the operator product expansion.

\end{document}